\documentclass[bibyear]{aa} % for the abstract without structuration 
\usepackage{graphicx}
\usepackage{epsfig}
\usepackage{natbib}
\usepackage{rotating}
\usepackage{tabularx}
\usepackage{subfig}
\bibpunct{(}{)}{;}{a}{}{,}
\usepackage{txfonts}

\usepackage{layouts}
\usepackage{amssymb}
\usepackage{upgreek}
\usepackage{amsmath}
\usepackage{gensymb}

\begin{document}

\title{NELIOTA: First temperature measurement of lunar impact
  flashes%\thanks{Based on observations made with the 1.2-m Kryoneri
%    telescope, operated by the National Observatory of Athens (NOA),
 %   Greece.}%\fnmsep\thanks{NELIOTA: NEO Lunar Impacts and Optical
%  TrAnsients}
}

\author{A.Z. Bonanos\inst{\ref{inst1}}, C. Avdellidou\inst{\ref{inst2}},
  A. Liakos\inst{\ref{inst1}}, E.M. Xilouris\inst{\ref{inst1}},
  A. Dapergolas\inst{\ref{inst1}},
  D. Koschny\inst{\ref{inst2},\ref{inst3}},\\ I. Bellas-Velidis\inst{\ref{inst1}},
  P. Boumis\inst{\ref{inst1}},
  V. Charmandaris\inst{\ref{inst1},\ref{inst4}},
  A. Fytsilis\inst{\ref{inst1}}, A. Maroussis\inst{\ref{inst1}}}

\institute{IAASARS, National Observatory of Athens, 15236 Penteli,
  Greece\label{inst1} \and Scientific Support Office, Directorate of
  Science, European Space Research and Technology Centre
  (ESA/ESTEC),\\ 2201 AZ Noordwijk, The Netherlands\label{inst2} \and
  Chair of Astronautics, Technical University of Munich, 85748 Garching,
  Germany\label{inst3} \and Department of Physics, University of Crete,
  71003, Heraklion, Greece\label{inst4}}
%\offprints{A.Z. Bonanos, \email{bonanos@noa.gr}}
\date{Received October 16, 2017; Accepted January 2, 2018}

\authorrunning{Bonanos et al.}

\titlerunning{NELIOTA: First temperatures of lunar impact flashes}

\abstract{We report the first scientific results from the NELIOTA (NEO
  Lunar Impacts and Optical TrAnsients) project, which has recently
  begun lunar monitoring observations with the 1.2-m Kryoneri
  telescope. NELIOTA aims to detect faint impact flashes produced by
  near-Earth meteoroids and asteroids and thereby help constrain the
  size-frequency distribution of near-Earth objects in the decimeter to
  meter range. The NELIOTA setup, consisting of two fast-frame cameras
  observing simultaneously in the $R$ and $I$ bands, enables -- for the
  first time -- direct analytical calculation of the flash
  temperatures. We present the first ten flashes detected, for which we
  find temperatures in the range $\sim1,600-3,100$~K, in agreement with
  theoretical values. Two of these flashes were detected on multiple
  frames in both filters and therefore yield the first measurements of
  the temperature drop for lunar flashes. In addition, we compute the
  impactor masses, which range between $\sim100$~g and $\sim50$ kg.}

\keywords{Moon - Meteoroids - Surveys}

\maketitle
%%%%%%%%%%%%%%%%%%%%%%%%%%%%%%%%%%%%%%%%%%%%%%%%%%%%%%%%%%%%%%%%%%%%%%%%%%%%%
%%%%%%%%%%%%%%%%%%%%%%%%%%%%%%%%%%%%%%%%%%%%%%%%%%%%%%%%%%%%%%%%%%%%%%%%%%%%%
\section{Introduction}

The estimation of the impact flux of near-Earth objects (NEOs) is
important not only for the protection of human civilization, but also
for the protection of space assets, which could be damaged or at least
perturbed even by small, mm- to cm-size impactors. A recent example that
reveals this necessity was the Chelyabinsk event in 2013, which was
caused by a 20-m impactor and member of the NEO population. This collision
was responsible for 1,500 injured civilians and a few thousand damaged
human assets in the area \citep{popova2013}. The precise flux density of
objects in this size range is not yet well known. A promising method for
constraining NEO flux densities in this size range is via the detection
of impact flashes on the Moon.
%Monitoring and detecting impact flashes on the Moon can help constrain
%the NEO number densities, as well as the size distribution of the
%population.  The observation of lunar impact flashes can contribute to
%better constraining the flux densities in this size range.

A plethora of laboratory impact experiments have been conducted over the
last 40 years initiated primarily to study spacecraft shielding, using
mainly metallic materials \citep[and references
  therein]{2002holsappleast3}. Apart from these technical experiments,
hypervelocity impacts (impact speeds $v>1$~km~s$^{-1}$) are also studied
at small scales. One important goal is to extrapolate the results to
larger size and velocity scales, towards the understanding of the
collisions on planetary surfaces by asteroids and comets or even among
the small bodies, for example the inter-asteroid collisions in the Main
Belt. It has been clearly shown that several impactor parameters, such
as the impactor's size, density, velocity and impact angle affect the
collision outcome, for example,\ the crater formation and the size and speed of
the ejecta plume
\citep{ryan1998,1999housen,2003housen,2011housen,me2016,me2017}. Observations
have shown that, for example, highly porous objects tend to be destroyed
tens of km above the surface of the Earth, as was the case with
2008~TC$_3$ \citep{2009jenniskens, 2010MAPS...45.1638B} and the Benesov
bolide \citep{borovicka1998}.

Telescopic surveys, such as the Catalina Sky Survey \citep{drake2009} and
Pan-STARRS \citep{chambers2016}, are continuously discovering new
objects, which are verified by follow-up observations from observers all
around the world.
%(https://cneos.jpl.nasa.gov/stats/site_all.html).
Space missions such as WISE and the {\it Spitzer} Space Telescope, along
with spectroscopic observations, provide valuable data to start
characterizing the physical properties of NEOs, such as their diameters,
albedos, and spectral types. Over 16,300 NEOs have been identified
\citep[as of August 2017]{jpl} of which only 1,142 have known diameters
$d$ \citep{delbo2017}, with the smallest ones being less than ten
meters. The NEO population consists of small bodies delivered from the
source regions of the Main Belt via mean motion and secular resonances
with the planets \citep{bottke2000,bottke2002}. Currently, the
completeness of the observed sample is at $d\sim1$~km, as surveys are
not able to massively detect NEOs that are smaller than a few tens of
meters in diameter. In fact, the very small bodies are usually detected
when their position on their orbit comes at close proximity to
Earth. For example, the 4-m near-Earth asteroid 2008~TC$_3$ was
discovered only 19~h prior to its impact on Earth and immediate radar
observations provided its size \citep{2009jenniskens}.

During the last decades, advances have been made by several groups
leading to a better estimation of the sizes of small impactors and their
flux on the Moon, correcting for the Earth as a target. This was done by
calculating the luminous efficiency $\eta$ of the detected flashes,
which is defined as the fraction of the impactor's kinetic energy ($KE$)
that is emitted as light at visible wavelengths ($L$), that is,\ $L = \eta
\times KE$. Great uncertainties occur during these calculations when the
events originate from sporadic NEOs (those not associated to meteor
streams), since the collisional velocity of the meteoroids on the Moon
is unknown. Several authors adopt average impact speeds for the lunar
surface spanning a wide range between 16 and 24~km~s$^{-1}$
\citep{ortiz2000,ortiz2002,suggs2014}. Uncertainties in the speed
estimation lead to uncertainties of the luminous efficiency value
$\eta$. The current estimations of the luminous efficiency of the lunar
impactors range over an order of magnitude resulting in weakly
constrained masses. However, when the impact events are linked to a
known meteoroid stream, this unknown parameter can be constrained
\citep[e.g.,][]{bellotrubio2000A} yielding masses that can be appended to
the current known impactors' size distribution \citep{harris_ast4} and
can also be used for further studies.

Apart from the NEO flux and size distribution, the lunar surface serves
as a large-scale impact laboratory to study the impact events. The term ``large-scale'' refers both to the impactor sizes and
speeds when comparing them to laboratory-based hypervelocity
experiments, where the sizes of impactors are typically a few mm and the
speeds below 8~km s$^{-1}$ \citep[e.g.,][]{1999MeScT..10...41B}. The
collisions of NEOs on the Moon give rise to several phenomena that can
be detected and further studied, such as impact cratering
\citep{speyerer2016}, seismic waves \citep{olberst1991} and the
enhancement of the lunar atmosphere with sodium
\citep{verani1998,smith1999}.

The light flash produced by an impact depends on several parameters,
including the mass and speed of the impactor. Even when the mass and
speed are known, the different combinations of mineralogical
compositions of both the target and impactor will affect the
result. Pioneering laboratory experiments were conducted more than 40
years ago, using dust accelerators and photomultipliers with filters at
several wavelengths, allowing the estimation of the plasma temperature
\citep{eichhorn1975,eichhorn1976, burchell1996B,
  burchell1996A}. Therefore the study of impact flashes could provide
insight to the complex problem of energy partitioning during an impact
event, when the majority of physical parameters are constrained or
measured (e.g.,\ mass and speed of the impactor, crater size, ejecta
speed).

The NELIOTA project\footnote{https://neliota.astro.noa.gr} (Xilouris et
al., in prep.) provides the first lunar impact flash observations
performed simultaneously in more than one wavelength band. In Section~2
we present the instrumentation, observation strategy and the first ten
lunar impact flashes from NELIOTA, providing their durations and
magnitudes. In Section~3, we focus on the first ever measurement of
impact flash temperatures using our two-colour observation technique,
while in Section~4 we present a new approach to estimate the impactors'
masses. The discussion and conclusions are given in Section~5.

\section{Observations}
\label{observations}

%\subsection{Instrumentation and Observation strategy}

NELIOTA has upgraded the 1.2-m Kryoneri
telescope\footnote{http://kryoneri.astro.noa.gr/} and converted it to a
prime-focus instrument with a focal ratio of f/2.8 for lunar monitoring
observations. The telescope has been equipped with two identical Andor
Zyla scientific CMOS cameras, which are installed at the prime focus and
are thermoelectrically cooled to 0 degrees. A dichroic beam-splitter
with a cut-off at 730~nm directs the light onto the two cameras
($2560\times2160$~pixels$^2$, $6.48~\mu$m per pixel), which observe in
visible and near-infrared wavelengths using $R$ and $I$ Cousin filters,
respectively. The maximum transmittance of each filter is at
$\lambda_{R}=641$~nm and $\lambda_{I}=798$~nm, corresponding to a
maximum quantum efficiency of $\sim50\%$ and $\sim40\%$,
respectively. The field-of-view of this setup is $16.0' \times
14.4'$. We use the $2\times2$ binning mode, which yields a pixel-scale
of $0.8''$, as it best matches the $1.2-1.5''$ average seeing and
results in a lower volume of data. Currently, the NELIOTA system has the
largest telescope with the most evolved configuration that performs
dedicated monitoring of the Moon, in search of faint lunar impact
flashes.

Observations are conducted on the dark side of the Moon, between lunar
phases $\sim$0.1 and 0.4. The maximum lunar phase during which
observations can be obtained is set by the strength of the glare coming
from the sun-lit side of the Moon. The observations begin/end
$\sim$20~min after/before the sunset/sunrise and last for as long as the
Moon is above an altitude of 20$\degr$. The altitude limit is set due to
limitations from the dome slit. The cameras simultaneously record at a
frame rate of 30 frames-per-second or every 33~ms, in $2\times2$ binning
mode. The exposure time of each frame is 23~ms, followed by a read-out
time of 10~ms. The observations are split into ``chunks'' that are
$15$~min in duration. At the end of each chunk, a standard star is
observed for calibration purposes. The standard stars have been
carefully selected a) to be as close as possible to the altitude of the
Moon and b) to have similar color indices to the expected colors of the
flashes (i.e.,\ $0.3<R-I<1.5$~mag). Flat-field images are taken on the
sky before or after the lunar observations, while dark frames are
obtained directly after the end of the observations. The duration of the
observations varies between $\sim$25~min and $\sim$4.5~h, depending on
the lunar phase and the time of year.

The novelty of the NELIOTA instrumentation setup is that it
simultaneously acquires data from two detectors and at two different
wavelengths. This setup enables the validation of a flash from a single
telescope and site, since a real event that is bright enough will be
detected by both cameras at the same position and at the same time,
whereas cosmic ray artefacts will only be detected by one camera at any
given position and time. Although satellites are also common artefacts,
they are typically recorded as streaks. Satellites moving at low enough
speeds so as not to show up as streaks in our 23-ms exposure time have
to be far away -- assuming, in the worst case, an object with a perigee
of 300~km, the apogee has to be at least at 17,500~km to result in an
apparent movement of less than 1 pixel per 23~ms. At this distance, an
object with a reflectivity of 0.5 would need to be at least 2~m in size
to be detected as a magnitude-11 flash. Geosynchronous satellites could
also produce artefacts due to reflection of sunlight off their solar
panels. However, since their positions are well known and clustered
around a declination of approximately zero, they pose no major concern
as artefacts, as they can be ruled out by using available catalogs of
geosynchronous satellite positions.

This paper presents and analyzes the first ten flashes that were
validated during the testing phase and the first months of the NELIOTA
campaign, from February to July 2017. These flashes originate from
sporadic NEOs. We checked various orbital catalogs of satellites and
could not find any objects in front of the Moon at the times of the
detected flashes. We note that the synchronization of the cameras during the
frame acquisition for these flashes is better than 6~ms. All validated
flashes are made available on the NELIOTA website within 24 hours of the
observations.

\subsection*{Photometry}
The data reduction is performed automatically by the NELIOTA pipeline
(described in Xilouris et al., in prep.) using the median images of the
respective calibration files (flat and dark images). These master-images
are used to calibrate the data of the Moon as well as those of the
standard stars. The pipeline searches for flashes on the images, after
computing and subtracting a running, weighted average image, which
removes the lunar background.

Due to the nonuniform background around a flash, which is caused by
surface features of the Moon (e.g.,\ craters, maria) and earthshine, we
performed photometry of the flashes on background-subtracted images. We
created these images by subtracting a median lunar image based on the five
frames before and five frames after the event. Aperture photometry with the
AIP4WIN software \citep{Berry00} was then performed for both the flashes
and standard stars observed nearest in time for each flash. Optimal
apertures corresponding to the maximum in the signal-to-noise ratio (S/N) of the
flux measurement were used for the flashes to avoid adding noise from
the subtracted background, while large apertures were used for the
standards. Since the standard stars are observed at approximately the
same airmass as the lunar surface, we can compute the flash magnitudes
in each filter as:
\begin{equation}
 m_{flash} =  m_{star} + 2.5 \log \left(\frac{S} {F}\right),
 \label{magnitudes}
\end{equation}
where m$_{star}$ and m$_{flash}$ are the calibrated magnitude of the
standard star and the magnitude of the flash, respectively, and $S$ and $F$
are the fluxes of the star and flash for the same integration time. All
photometric measurements and error determinations were independently
computed using the IRAF\footnote{IRAF is distributed by the National
  Optical Astronomy Observatory, which is operated by the Association of
  Universities for Research in Astronomy (AURA) under cooperative
  agreement with the National Science Foundation.}  \textit{apphot}
package and were found to agree within errors with the results from
AIP4WIN.

Table~\ref{table1} presents the date and universal time at the start of
the observation for each impact flash detection, its $R$ and $I-$band
magnitude and error, the duration recorded in $I$, as well as the
temperature and mass measurements are described in the following Sections. The
flash durations are estimated by multiplying the 33-ms frame rate by the
number of frames the flash was detected on and are thus upper limits to
the real flash duration. The durations range between 33 and 165 ms, in
agreement with previously reported values
\citep[e.g.,][]{Yanagisawa02}. We note that Flashes 2, 6, 7, and 10 were
detected over multiple, consecutive frames. Flashes 2 and 10 are the
brightest flashes in the current dataset and had simultaneous detections
in both bands in consecutive frames. They are used below to measure the
temperature evolution of the flashes.

\begin{table*}%[!Ht]
\centering
\caption{Dates, universal times (UT), magnitudes in each filter,
  duration recorded in the $I-$band and listed for the first entry of
  each flash, temperatures and impactor masses of the first ten NELIOTA
  flashes. Multiple entries per flash correspond to the consecutive
  frames they were detected on. Masses are calculated for both
  $\eta_{1}$ and $\eta_{2}$ values (see text for
  details).}\label{table1}
%\begin{tabular}{l|ccrrclrr}
\begin{tabular}{l|ccrrclrr}
\hline\hline
Flash & Date & UT& R $\pm$ $\sigma_R$ & I $\pm$ $\sigma_I$ & Duration & T $\pm$ $\sigma_T$ & Mass ($\eta_1$) $\pm$ $\sigma_M$& Mass ($\eta_2$) $\pm$ $\sigma_M$\\
&&& (mag) & (mag) & (ms) & (K) & (kg) & (kg) \\
\hline
1\tablefootmark{a} & 2017-02-01 & 17:13:57.863 &10.15 $\pm$ 0.12 & 9.05 $\pm$ 0.05 & 33 & 2350 $\pm$ 140 & 0.6 $\pm$ 0.3 & 0.2 $\pm$ 0.1\\
\hline
2\_1\tablefootmark{b} & 2017-03-01 & 17:08:46.573 & 6.67 $\pm$ 0.07 & 6.07 $\pm$ 0.06 & 132 & 3100 $\pm$ 30 & 4.4 $\pm$ 0.5 & 1.6 $\pm$ 0.2\\
2\_2 & 2017-03-01 & 17:08:46.606 & 10.01 $\pm$ 0.17 & 8.26 $\pm$ 0.07 & $-$ & 1775 $\pm$ 100 & $-$ & $-$\\
2\_3 & 2017-03-01 & 17:08:46.639 & $-$ & 9.27 $\pm$ 0.10 & $-$ & $-$ & $-$ & $-$\\
2\_4 & 2017-03-01 & 17:08:46.672 & $-$ & 10.57 $\pm$ 0.15 & $-$ & $-$ & $-$ & $-$\\
\hline
3\tablefootmark{b} & 2017-03-01 & 17:13:17.360 & 9.15 $\pm$ 0.11 & 8.23 $\pm$ 0.07 & 33 & 2568 $\pm$ 130 & 0.9 $\pm$ 0.4 & 0.4 $\pm$ 0.1\\
\hline
4\tablefootmark{c} & 2017-03-04 & 20:51:31.853 & 9.50 $\pm$ 0.14 & 8.79 $\pm$ 0.06 & 33 & 2900 $\pm$ 270 & 0.4 $\pm$ 0.3 & 0.2 $\pm$ 0.1\\
\hline
5\tablefootmark{d} & 2017-04-01 & 19:45:51.650 &10.18 $\pm$ 0.13 & 8.61 $\pm$ 0.03 & 33 & 1910 $\pm$ 100 & 2.3 $\pm$ 1.0 & 0.8 $\pm$ 0.4\\
\hline
6\_1\tablefootmark{e} & 2017-05-01 & 20:30:58.137 &10.19 $\pm$ 0.18 & 8.84 $\pm$ 0.05 & 66 & 2070 $\pm$ 170 & 1.3 $\pm$ 0.9 & 0.5 $\pm$ 0.3\\
6\_2 & 2017-05-01 & 20:30:58.170 & $-$ & 10.44 $\pm$ 0.21 & $-$ & $-$ & $-$ & $-$\\
\hline
7\_1\tablefootmark{f} & 2017-06-27 & 18:58:26.680 &11.07 $\pm$ 0.32 & 9.27 $\pm$ 0.06 & 66 & 1730 $\pm$ 210 & 2.5 $\pm$ 2.4 & 0.9 $\pm$ 0.9\\
7\_2 & 2017-06-27 & 18:58:26.713 & $-$ & 10.80 $\pm$ 0.21 & $-$ & $-$ & $-$ & $-$\\
\hline
8\tablefootmark{g} & 2017-07-28 & 18:42:58:027 &10.72 $\pm$ 0.24 & 9.63 $\pm$ 0.10 & 33 & 2340 $\pm$ 310 & 0.4 $\pm$ 0.5 & 0.2 $\pm$ 0.2\\ 
\hline
9\tablefootmark{g} & 2017-07-28 & 18:51:41.683 &10.84 $\pm$ 0.24 & 9.81 $\pm$ 0.09 & 33 & 2410 $\pm$ 310 & 0.3 $\pm$ 0.3 & 0.1 $\pm$ 0.1\\  
\hline
10\_1\tablefootmark{g} & 2017-07-28 & 19:17:18.307 & 8.27 $\pm$ 0.04 & 6.32 $\pm$ 0.01 & 165 & 1640 $\pm$ 20 &55 $\pm$ 19 & 20 $\pm$ 7\\
10\_2 & 2017-07-28 & 19:17:18.340 & 9.43 $\pm$ 0.12 & 7.44 $\pm$ 0.02 & $-$ & 1620 $\pm$ 70 & $-$ & $-$\\
10\_3 & 2017-07-28 & 19:17:18.373 & $-$ & 8.89 $\pm$ 0.07 & $-$ & $-$ & $-$& $-$\\
10\_4 & 2017-07-28 & 19:17:18.406 & $-$ & 9.38 $\pm$ 0.11 & $-$ & $-$ & $-$& $-$\\
10\_5 & 2017-07-28 & 19:17:18.439 & $-$ & 10.29 $\pm$ 0.23 & $-$ & $-$ & $-$& $-$\\
\hline
\end{tabular}
\tablefoot{The standard stars used for the calibration of each flash
  are: \tablefoottext{a}{SA 92$-$263,} \tablefoottext{b}{SA 93$-$333,}
  \tablefoottext{c}{SA 97$-$345,} \tablefoottext{d}{LHS 1858,}
  \tablefoottext{e}{2MASS J09212193$+$0247282,} \tablefoottext{f}{GSC
    04932$-$00246,} \tablefoottext{g}{GSC 00362$-$00266.}}
\end{table*}

%%%%%%%%%%%%%%%%%%%%%%%%%%%%%%%%%%%%%%%%%%%%%%%%%%%%%%%%%%%%%%%%%%%%%%%%%%%%%
%%%%%%%%%%%%%%%%%%%%%%%%%%%%%%%%%%%%%%%%%%%%%%%%%%%%%%%%%%%%%%%%%%%%%%%%%%%%%
\section{Temperature estimation of the impact flashes}
\label{temperature}
The NELIOTA observations provide the first observational evidence for
the temperature of impact flashes. Since we measure the emitted flux
density in two different filters ($R$ and $I$), we can determine the
flash temperature by comparing the intensities in the two wavelength
bands. Assuming black-body emission %(i.e. the emission follows a Planck
%curve)
\citep{eichhorn1975, burchell1996A,ernst2004,suggs2014}, a given
temperature will result in a specific ratio between the measured
intensities in the $R$ and $I-$bands.

The ratio of the energies $E_{1}/E_{2}$ released in two different
wavelengths depends only on the temperature $T$. Here we present an
analytical method for calculating the temperatures of the NELIOTA
flashes. The Planck formula is given by:
\begin{equation}
B(\lambda,T) = \frac{2hc^{2}}{\lambda^{5}}\frac{1}{\exp(\frac{hc}{\lambda k_{B}T}) - 1},
\end{equation}
where $h=6.62 \times 10^{-34}$\;kg\;m$^{2}$\;s$^{-1}$ is the Planck
constant, $c=3 \times 10^{8}$\;m\;s$^{-1}$ the speed of light,
$k_{B}$\;=\;1.38$\times$10$^{-23}$\;kg\;m$^{2}$\;s$^{-2}$\;K$^{-1}$ the
Boltzmann constant, $T$ and $\lambda$ the temperature of the flash and
the wavelength of the photons, respectively. Dividing the Planck formula
with the energy $E=h c/\lambda$ per photon, we obtain the photon
radiance per wavelength $L_P(\lambda,T)$:
%The total energy that we measure from an impact flash is the sum of the
%energies of the $N$ individual photons detected by the camera system,
%$E_{total}$~=~$\frac{Nhc}{\lambda}$, where $E$~=~$h \nu$ is the energy
%of a photon having frequency $\nu$~=~$c/\lambda$. The two unknown
%parameters in Eq.~2 are the energy $E$ and the temperature $T$.  By
%replacing the received energy with the number of photons $N$ (assuming
%an effective wavelength) we obtain the photon radiance per wavelength
%$L_P(\lambda,T)=B(\lambda,T)/hc/\lambda$:

\begin{equation}
L_P(\lambda,T) = \frac{2c}{\lambda^{4}}\frac{1}{\exp(\frac{hc}{\lambda k_{B}T}) - 1}
.\end{equation}
Equation~3 is now linked to the absolute flux, $f_{\lambda}$, of the
flash as:
\begin{equation}
f_{R} = \Omega L_P(R,T) \quad\text{and}\quad f_{I} = \Omega L_P(I,T) \quad\text{for each filter,}\\
\end{equation}
where $\Omega$ is a constant. Since the observations are performed
simultaneously at two different wavelengths, $R$ and $I$, we measure the
two instrumental fluxes for the flash ($F_{R}$ and $F_{I}$) and for the
standard star ($S_{R}$ and $S_{I}$). These measured fluxes are linked to
the absolute ones ($f_{R},f_{I}$ and $s_{R},s_{I}$) with the factors
$\xi_{R}$ and $\xi_{I}$, which depend on the instrument and atmospheric
transmission. Therefore, for each $\lambda$ we get:
\begin{subequations}
\begin{equation}
F_{R} = \xi_{R} f_{R} \quad\text{and}\quad  F_{I} = \xi_{I} f_{I}\quad \text{for the flash,}
\end{equation}
\begin{equation}
S_{R} = \xi_{R} s_{R}  \quad\text{and}\quad S_{I} = \xi_{I} s_{I}\quad\text{for the star.}
\end{equation}
\end{subequations}

Using the color of the standard star ($R-I$), which is known from the literature, and the ratio of Eq.~5b we obtain the value of the ratio of $\xi_{I}/\xi_{R}$,

\begin{subequations}
\begin{equation}
R-I= -2.5 \log \left(\frac{s_{R}}{s_{I}}\right) = -2.5 \log \left(\frac {\xi_{I}} {\xi_{R}}\frac{S_{R}}{S_{I}}\right)
,\end{equation}
\begin{equation}
\xi=\frac {\xi_{I}} {\xi_{R}} = \frac{S_{I}}{S_{R}} 10^{-0.4 \;(R-I)}.
\end{equation}
\end{subequations}

\begin{figure}[!hb]
 \includegraphics[width=0.35 \textwidth, angle=270]{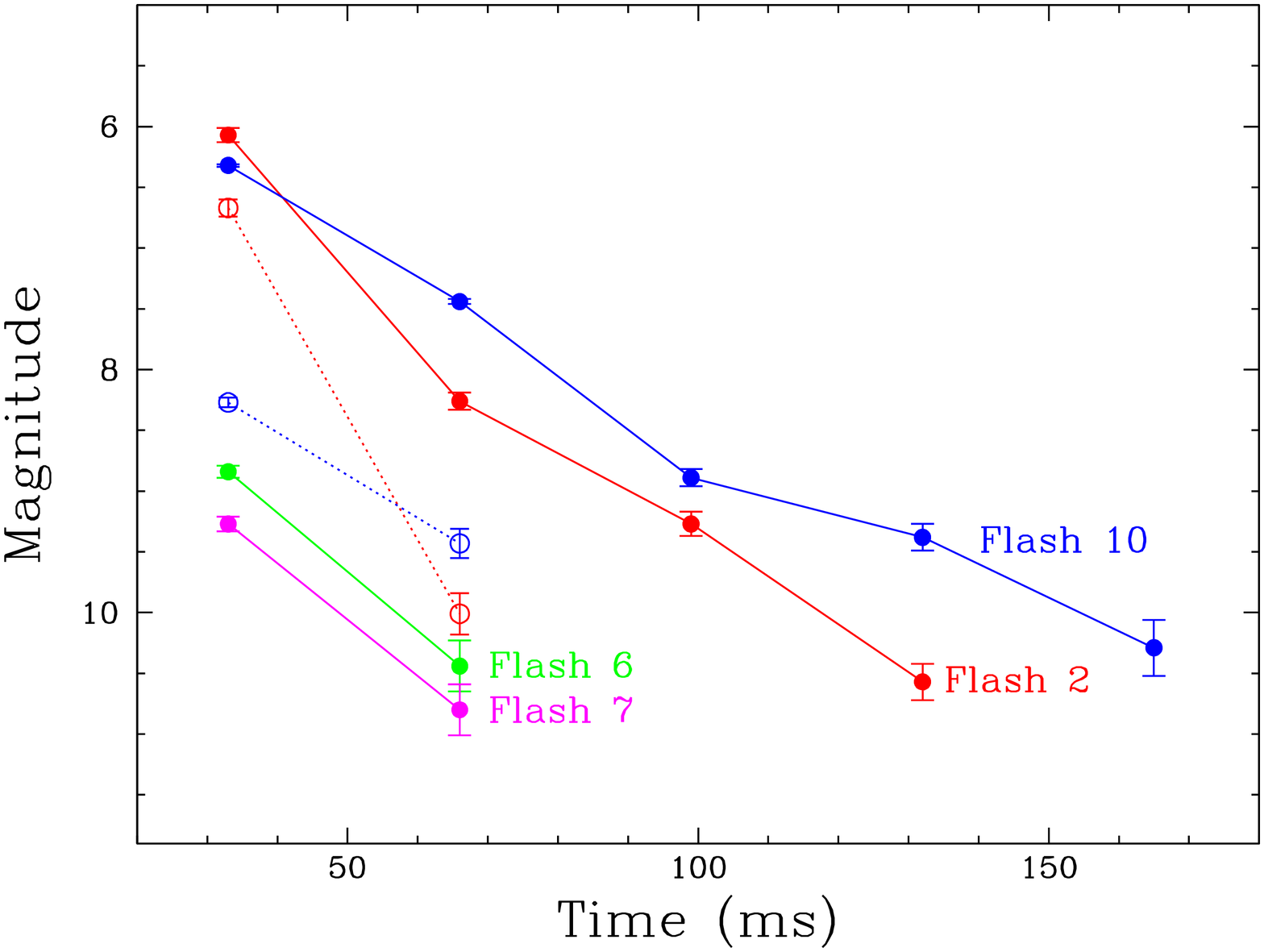}
 \includegraphics[width=0.35 \textwidth, angle=270]{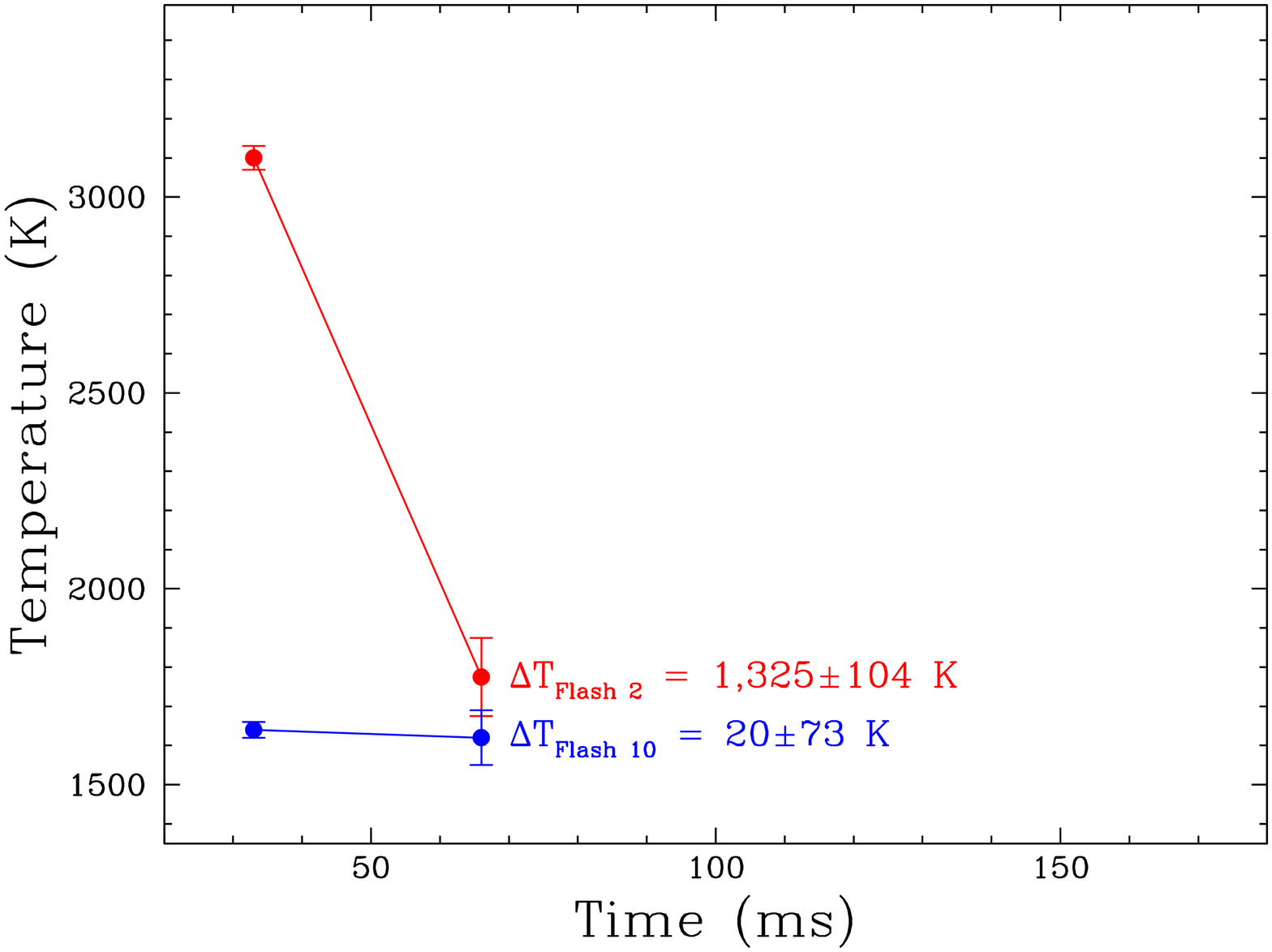}
  \caption{\textit{Upper panel:} Light curves of the four multi-frame
    events in the $I$ (filled circles; solid line) and $R$ (open
    circles; dotted line) bands. \textit{Lower panel:} Temperature
    evolution for Flashes 2 and 10.}
\label{flash}
\end{figure}

The $\xi$ value is now used to find the ratio of the flash flux in both
filters $f_{R}/f_{I}$ using Eq.~5a.  From the ratio of Eq.~4,
substituting the $L_P(R,T)$/$L_P(I,T)$ expressions from Eq.~3 and the
$f_{R}/f_{I}$ using Eq.~5a, we have:
\begin{equation}
\frac {L_P(R,T)}{L_P(I,T)} = \xi \frac {F_{R}} {F_{I}}  
,\end{equation}
 and thus the temperature $T$ becomes the only unknown parameter, which
 is calculated numerically using Eq.~7.

For each event, we performed 10$^{5}$ Monte Carlo simulations in order
to compute the standard deviation of each temperature measurement. At
each iteration, random numbers were obtained from the observed flux
distribution. The values of $f_{R}$ and $f_{I}$ were extracted from a
Gaussian distribution centered at the nominal value of each flux, while
adopting the standard deviation that resulted from the photometry. All
temperatures and their uncertainties are presented in
Table~\ref{table1}.

The multi-frame Flashes 2 and 10 enable us to calculate the drop of the
temperature for the first time, as they have simultaneous detections in
both bands in consecutive frames. We find a temperature decrease of
1,325~$\pm$~104~K for Flash 2 and 20~$\pm$~73~K for Flash 10,
that is,\ between the first detection and the subsequent one 33~ms
later. The temperature evolution appears very different for each case
and indicates a large difference in the impactor size, as a larger and
heavier object will take longer to cool. A larger sample of multi-frame
flashes from NELIOTA will allow us to determine the cooling behavior of
the flashes and its relation to the impactor mass.  Figure~\ref{flash}
illustrates the light curve evolution for the four multi-frame flashes
and temperature evolution for Flashes 2 and 10. The data are plotted at
the end of the frame read-out of the corresponding measurement. All
$I-$band light curves have a similar slope. Flash 2 presents a steeper
decrease in the $R-$band than in the $I-$band.

%%%%%%%%%%%%%%%%%%%%%%%%%%%%%%%%%%%%%%%%%%%%%%%%%%%%%%%%%%%%%%%%%%%%%%%%%%%%%
%%%%%%%%%%%%%%%%%%%%%%%%%%%%%%%%%%%%%%%%%%%%%%%%%%%%%%%%%%%%%%%%%%%%%%%%%%%%%
\section{Mass estimation of the impactors}
\label{mass}
The first step for the mass estimation is to derive the luminosity $L$
of the impact event. Given that observations up to now were mostly
carried out using a single $R-$band filter, the value of $L$ was not
well constrained \citep{bellotrubio2000A, bouley2012, ortiz2015,
  madiedo2015, suggs2014, suggs2017}. In this paper, we are able to
estimate $T$ for the first time from the two wavelength bands provided
by NELIOTA, and therefore can directly derive the luminous
energy. Assuming black-body radiation from a spherical area, the
bolometric energy is expressed in Joules as:
\begin{equation}
L = \sigma A T^4 t,
\end{equation}
where $\sigma=5.67 \times 10^{-8}$\;W\;m$^{-2}$\;K$^{-4}$ is the
Stefan-Boltzmann constant, $A=2 \pi r^{2}$ the emitting area of radius
$r$ for a flash near the lunar surface, $T$ the flash temperature
derived above, and $t$ the exposure time of the frame when the photons
were integrated. However, this calculation is not straightforward since
we do not know the size of the radiating plume. A reasonable assumption
is that the flashes are not resolved and thus the area is smaller than
the pixel-scale ($0.8"$), which corresponds to a linear distance of
$\sim$1,500~m at the center of the Moon's disk.

The flux of the event at a specific wavelength, $f_{\lambda}$, is
related to Planck's law expressed in photon radiance per wavelength (as
described in Eq.~3 \& 4):
\begin{equation}
f_{\lambda} = \frac{L_P(\lambda, T)\epsilon \pi r^{2}}{D^{2}},
\end{equation}
where $r$ is the radius of the radiative area, $D$ the Earth-Moon
distance at the time of the observation and $\epsilon$ the emissivity,
which we assume to be 1. The monochromatic flux of the flash
$f_{\lambda}$ can be calculated from:
\begin{equation}
m_{\lambda}-m_{Vega(=0)} = -2.5 \log \left(\frac{f_{\lambda}}{f_{Vega,\lambda}}\right)
;\end{equation}
therefore Eq.~9 can be solved for the unknown area of radius $r$. For
the error estimation in $r$, we followed the approach described for the
$T$ error estimation. We performed Monte Carlo simulations for the
absolute flux estimation using Eq.~10, by randomly selecting flash
magnitudes ($m_{\lambda}$) from their Gaussian distribution, with
centers and standard deviations from the values of
Table~\ref{table1}. This procedure was repeated for each filter and
returned the absolute fluxes with their 1$\sigma$ values. In turn, these
values were used as input for new Monte Carlo simulations, the
calculation of $r$, and the final value comes from the average of the
$A$-value that was found for each filter. We use a simple average of the
two derived areas (one for each filter) for a single event since the
differences were small. The luminosity of the flash $L$, which now can
be easily derived from Eq.~8, is just a fraction $\eta$ (luminous
efficiency) of the impactor's initial kinetic energy $KE$:
\begin{equation}
KE = \frac{L}  {\eta}= \frac{1}{2} m v^2,
\end{equation}
where $m$ in kg is the mass of the impactor and $v$ in m~s$^{-1}$ the
impact speed. In this work we use the formula derived by
\citet{swift2011} and also used by \citet{suggs2014}:
\begin{equation}
\eta = 1.5\times 10^{-3} e^{-(v_o/v)^2 } ,\end{equation} where
$v_o=9.3$~km s$^{-1}$, in order to estimate the luminous efficiencies
$\eta_{1}$ and $\eta_{2}$ for two extreme impact velocities,
16~km~s$^{-1}$ and 24~km~s$^{-1}$ \citep{steel1996,mcnamara2004},
respectively. Table~\ref{table1} presents the resulting masses, which
range between $0.3-55$~kg for $\eta_{1}=1.07 \times 10^{-3}$ and
$0.1-20$~kg for $\eta_{2}=1.29 \times 10^{-3}$.

%%%%%%%%%%%%%%%%%%%%%%%%%%%%%%%%%%%%%%%%%%%%%%%%%%%%%%%%%%%%%%%%%%%%%%%%%%%%%
%%%%%%%%%%%%%%%%%%%%%%%%%%%%%%%%%%%%%%%%%%%%%%%%%%%%%%%%%%%%%%%%%%%%%%%%%%%%%
\section{Discussion and Conclusions}

NELIOTA is the first lunar monitoring system that enables the direct
temperature measurement of observed lunar impact flashes, thanks to its
unique twin camera and two-filter observation setup. Until now, the
temperature could only be estimated, as it was based on modeling or
experimental work.  For example, \citet{suggs2014} used $T$=2,800~K from
\citet{nemtchinov1998}. \citet{cintala1992} suggested that the flash
temperatures, which depend on the type of material on the lunar surface,
should range between 1,700~K and 3,800~K. The agreement of the values we
obtain for the first NELIOTA flashes ($\sim$1,600--3,100~K) with the
theoretical range is of great importance for estimating the luminosity
of an impact flash and therefore its mass and size.

The estimation of the masses of the impactors is a challenging procedure
because many factors contribute to the mass uncertainty. The
uncertainties of the observed fluxes contribute to the temperature
estimation, which propagates to the calculation of the radiating area
$A$ and then to the calculation of the bolometric luminosity
$L$. However, the parameter that has the most important effect for the
mass estimation is the luminous efficiency. Luminous efficiencies
derived from laboratory experiments \citep{ernst2005} tend to be smaller
by a few orders of magnitude compared to the ones derived from
observations.  Previous studies have proposed various values for the
luminous efficiency, for example,\ $\eta\sim 2 \times 10^{-3}$ from
observations of lunar Leonids \citep{bellotrubio2000A}. While we compute
$\eta$ from Eq.~12 to be $\sim1.1-1.3 \times 10^{-3}$, other extreme
values have been used for the sporadic impactor
population. Specifically, when values in the range 10$^{-3}$$<\eta<$
10$^{-4}$ are used, the mass of the same impactor can differ by an order
of magnitude, even larger than the one we calculate here. Since large
uncertainties exist in the calculation of the mass due to the unknown
impact velocity, any estimation of the size will also be
uncertain. Despite these uncertainies, our mass estimates (100 g to 55
kg) are at least an order of magnitude higher than the values (0.4 g to
3.5 kg) reported by \citet{suggs2014}.
 
The impactors can be either asteroidal or cometary in origin, implying a
difference in the density. Even if we consider the scenario that the
bodies are near-earth asteroids, their densities can span a large range.
Bulk densities of asteroids differ according to their mineralogy and
macroporosity \citep{2002BRITTast3, carry2012}. However, there now
exists a large collection of meteorites, pieces of asteroids, and an
advanced knowledge of their densities. Average bulk densities of
meteorites are between 1,600 and 7,370~kg~m$^{-3}$, where these extremes
correspond to carbonaceous and iron meteorites, respectively
\citep{consolmagno1998, britt2003, consolmagno2008, macke2010,
  macke2011a, macke2011b}. For all these reasons, new laboratory
experiments using several types of materials will be very important for
understanding impacts and the flash-generation mechanism, as they will
provide a database of the impact parameters and their correlations
(mass, impact speed, composition, flash duration, etc.).

In summary, we report the first ten lunar impact flashes detected by the
NELIOTA project, using the 1.2~m Kryoneri telescope. The multi-band
capability of the NELIOTA cameras enables us to directly measure the
temperatures of the impact flashes for the first time and to estimate
the impactor masses. We find the measured temperature values
($\sim$1,600--3,100~K) to agree with previously published theoretical
estimates, as discussed above. Furthermore, our sample contains four
multi-frame flashes, two of which offer the opportunity for the
estimation of the temperature evolution of the flash. We find a decrease
of $1,325\pm104$~K for Flash 2 in 33~ms, while the decrease in the same
time interval for Flash 10 ($20\pm73$~K) is consistent with zero. This
difference is likely related to the fact that the impactor producing
Flash 10 has a mass that is an order of magnitude larger than that of
the impactor producing Flash 2. We also note that Flash 10 does not
appear as a point source. We expect future detections of multi-frame and
multi-band flashes with NELIOTA to provide a large enough sample for it
to help determine the temperature evolution properties of impact
flashes. Furthermore, our mass estimations rely on direct measurement of
the luminous energy, given the directly measured temperature. The mass
estimates that we report (100 g to 55 kg) are higher than previous
estimations, despite the range of values assumed for the impact
velocities and the resulting values of $\eta$.

Obtaining NEO flux densities requires increasing the number of
measurements of lunar impact flashes made during meteor showers. These
will be important for estimating the impactor sizes, since their impact
velocity will be constrained. NELIOTA is expected to contribute to
detections of stream impact flashes, which will also constrain the
critical, yet uncertain, value of $\eta$. The multi-band capability of
NELIOTA will generate valuable statistics on the temperatures of impact
flashes and their evolution. The comparison of these measurements with
the laboratory results will provide insight to the physics of impact
flashes.

\begin{acknowledgements}
AL, EMX, AD, IBV, PB, AF and AM acknowledge financial support by the
European Space Agency under the NELIOTA program, contract
No. 4000112943. This work has made use of data from the European Space
Agency NELIOTA project, obtained with the 1.2-m Kryoneri telescope,
which is operated by the Institute for Astronomy, Astrophysics, Space
Applications and Remote Sensing, National Observatory of Athens,
Greece. Thanks to Danielle Moser, Robert Suggs and Steven Ehlert (NASA
Marshall Space Flight Center) for discussions and comments on the
manuscript. CA would like to thank Regina Rudawska and Elliot
Sefton-Nash (ESTEC/ESA) for input.
\end{acknowledgements}

\end{document}